\documentclass[journal]{IEEEtran}
\usepackage{booktabs}
\usepackage{epsfig}
\usepackage{subfigure}
\usepackage{float}
\usepackage{url}
\usepackage{amsfonts}
\usepackage{notoccite}
\usepackage{amsmath}
\usepackage{amssymb}
\usepackage{graphicx}
\usepackage{color}
\usepackage{multirow}
\usepackage{rotating}
\usepackage[ruled,vlined]{algorithm2e}

\begin{document}
\title{Learning to Charge RF-Energy Harvesting Devices in WiFi Networks}
%
%

\author{Yizhou Luo and Kwan-Wu Chin
\thanks{The authors are with the School of Electrical, Computer and Telecommunications Engineering, University of Wollongong, NSW, Australia.  Email: yl631@uowmail.edu.au, kwanwu@uow.edu.au}}

\maketitle

%
\begin{abstract}
In this paper, we consider a solar-powered Access Point (AP) that is tasked with supporting both non-energy harvesting or {\em legacy} data users such as laptops, and devices with Radio Frequency (RF)-energy harvesting and sensing capabilities. We propose two solutions that enable the AP to manage its harvested energy via transmit power control and also ensure devices perform sensing tasks frequently.  Advantageously, our solutions are suitable for current wireless networks and do not require perfect channel gain information or non-causal energy arrival at devices.  The first solution uses a deep Q-network (DQN) whilst the second solution uses Model Predictive Control (MPC) to control the AP's transmit power.
Our results show that our DQN and MPC solutions improve energy efficiency and user satisfaction by respectively 16\% to 35\%, and 10\% to 42\% as compared to competing algorithms.
\end{abstract}


\IEEEpeerreviewmaketitle

\section{Introduction}\label{intro}
%
%
The Internet of Things (IoT) network~\cite{IIoTbb} will play a critical role in our daily activities.  In particular, sensing devices will be an integral part of devices purchased by consumers.  For example, as detailed in \cite{WiWear}, wearable devices with sensing capabilities such as a thermometer, video camera or smart watch are now available commercially to enable smart homes/offices as well as for video surveillance. Another example is \cite{IoTLearn}, whereby a motion sensor is used to infer usage of appliances in a home.
Moreover, these devices are likely to harvest energy from Wireless Local Area Networks (WLANs), which are now ubiquitous and widely used to connect conventional IEEE 802.11 devices such as iPads~\cite{WiWear}.
Energy efficiency is also becoming a concern~\cite{IIoTaa}.  To this end, future  WLANs are likely to incorporate Energy Harvesting (EH) Access Points (APs) in order to reduce carbon emissions and operating expenditure~\cite{EHAP}. Moreover, IoT devices will have Radio Frequency (RF)-energy harvesting capability; see \cite{POWERIOT} for a prototype that uses transmissions from an AP to power an on-board camera.

\begin{figure}[htbp]
	\centering
	\centerline{\includegraphics[scale=0.60]{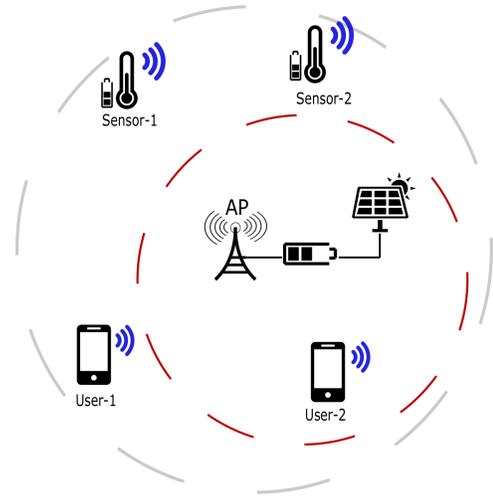}}
	\caption{An example network with a solar-powered AP and both legacy data users and RF-energy harvesting IoT devices. Both sensors receive energy whenever the AP uses a high transmit power level, denoted by the gray circle.}
	\label{IOTWLAN}
\end{figure}

Figure~\ref{IOTWLAN} shows a solar-powered AP that serves not only {\em legacy} data users/devices, which do not have RF-energy harvesting capability, but also nearby IoT devices equipped with a temperature sensor and an RF-energy harvester. All nodes operate on the {\em same} frequency. 
Whenever the AP delivers data to {\em legacy} users, IoT devices harvest RF-energy. The amount of harvested RF-energy is a function of their distance to the AP, time-varying channel gains, how often the AP transmits and also the AP's transmit power; Figure~\ref{IOTWLAN} shows two possible transmit power levels.  A key challenge here is that the transmit power used by the AP is dependent on its available energy, which exhibits spatio-temporal properties.  Critically, the AP only has {\em causal} knowledge of its solar energy arrivals; i.e., the AP only knows its current and past energy arrivals.
Another challenge is that IoT devices may be tasked with returning their sample data periodically. However, they may not have harvested sufficient energy from the AP's data transmissions. 

Henceforth, our aim is to determine a transmit power allocation policy that allows an AP to support the data rate of {\em legacy} data users, and at the same time delivers RF-energy to IoT devices.
In this respect, this paper contains the following contributions.
We present two solutions to derive the said policy and compare their performance in improving user satisfaction. The first solution relies on the Deep Q-Network (DQN) framework, where given an AP's state comprising of its energy level and legacy user channel gain, it determines the best transmit power control that yields the maximum satisfaction for both user types.
The second solution uses Model Predictive Control (MPC), where the AP uses Gaussian Process Regression (GPR), a machine learning method, to predict future harvested energy and channel gains.
Both solutions do not rely on perfect Channel State Information (CSI) to IoT devices and assume causal energy arrivals at the AP. These assumptions are made for practical reasons to ensure our solutions are readily deployable in current wireless networks.

Next, in Section~\ref{Related Works}, we identify gaps in past works.  Section~\ref{System} formalizes our system and problem.  After that Section~\ref{Solution} outlines two solutions and they are evaluated in Section~\ref{Evaluation}. Section~\ref{CONC} concludes the paper.

%
\section{Related Works}\label{Related Works}
%
%

Numerous works have considered RF-charging or Wireless Powered Communication Networks (WPCNs); see \cite{RF1} and references therein. They mainly focus on maximizing the sum-rate at an Hybrid Access Point (HAP)~\cite{6878442, 6678102}. However, they mostly consider optimizing charging and transmission slots, which is different to our problem. Our work overlaps with those that apply Reinforcement Learning (RL) to adjust the transmission power of an HAP. For example, in~\cite{7959919},  the goal is for a HAP to learn a robust transmission/changing power control policy to minimize the number of dropped packets due to attacks.
These works, however, only consider one type of users; namely RF-energy devices. Also, their HAP does not harvest energy. Critically, they assume the current channel condition of devices is known by HAPs. We do not make such assumptions.
Another group of works considers EH HAPs~\cite{8480633, 8843917} to charge RF-energy devices. They aim to transfer energy to RF devices in order to maximize the sum-rate or lifetime of devices. In~\cite{8480633}, the authors determine the transmission power of HAPs and control the active/sleep modes of HAPs. The aim is to maximize the lifetime of IoT devices given casual energy arrivals at a HAP. In \cite{8843917}, the authors propose to manage the energy consumption of HAPs for tasks such as sensing, computing, and uplink transmission. The aim is to minimize the average consumption of on-grid energy while ensuring IoT devices have sufficient energy or capacity. However, their system is different from ours because we consider RF changing in an existing WiFi systems.

To date, only a small number of works have considered supporting two types of users; namely, \cite{POWERIOT, 8215401, 8501909, 8809880}. 
In \cite{POWERIOT}, APs inject a power packet whenever their data queue length is shorter than a given number of packets.  Reference~\cite{8215401} aims to optimize the transmissions of APs in order to i) maximize the amount of energy delivered to an RF sensor, and ii) minimize the total number of packets queued at APs.  In \cite{8501909}, the authors study the impact of interference on both data users and RF-energy users. The authors aim to optimize sub-carrier allocation and transmission power control in order to minimize the total energy consumption of their system.
Reference~\cite{8809880}  aims to determine the beacon frequency of an AP and the charging period of IoT devices that maximize their lifetime. 
Our work is fundamentally different to these works.  First, the APs or base stations of these works have no EH capability. Second, our work aims to control the transmission power at an AP with stochastic energy arrivals in order to satisfy both RF-energy devices and legacy WiFi users.  Another distinction to prior works is that we consider imperfect CSI to IoT devices and causal knowledge of energy arrivals.
Moreover, they use mathematical optimization as a solution approach and requires non-causal information. On the other hand, we employ machine learning approaches and adapt to historical CSI and energy arrivals.

%
\section{System Model and Problem}\label{System}
We assume time is discrete, and the set of time slots is $\mathcal{T} = \{1, 2, \ldots, {T}\}$. Each slot is one second in length; this means the term power and energy can be used interchangeably. There are $N$ IoT devices uniformly located around an AP. Let $\mathbb {D}$ be the set of IoT devices. Similarly, there are $U$ legacy data users. In each time slot, the AP serves one data user.

The AP has a battery of size $B_{max}$. Its energy arrival is governed by the Markovian model presented in~\cite{7008488}. Specifically, the model contains four different solar states: `Excellent', `Good', `Fair', and `Poor'. Each state represents different solar intensity throughout a day. In the $j$-th state, the energy arrival $x$ (in mJ) is a random value drawn from a Normal distribution $\mathcal{N}(x\,|\,\mu_j,\sigma_j)$ with mean $\mu_j$ and variance $\sigma_j$. Then, the energy harvested by the AP in slot $t$ is $\widetilde{E}_t=x_t\Phi\overline{\eta}$, where $\Phi$ and $\overline{\eta}$ is the panel size and the solar energy conversion efficiency. The energy level of the AP, denoted as $B_t$, evolves as per,
\begin{equation}
    B_t=\min(B_{max},\,B_{t-1}-P_{t-1}+\widetilde{E}_t),
\end{equation}
where $P_t$ (in mW) is the transmission power of the AP at time slot $t$, which is bounded by $P_{max}$. 

We consider block fading. The AP is aware of the CSI to {\em legacy} or data users but is unaware of the CSI to IoT devices.
Let $g^t_{i}$ be the channel gain between the AP and IoT device $i$, and it is defined as,
\begin{equation}\label{channel gain}
    g^t_{i}=\frac{1}{{d_{i}}^2}{|Z|}^2,
\end{equation}
where $Z$ is drawn from a complex normal distribution $\mathcal{CN}(\mu, \sigma^2)$, and $d_{i}$ denotes the of device $i$ from the AP. 

IoT devices are equipped with a battery with capacity $b_{max}$. Let $b_i^t$ be the current battery level of IoT device $i$. In each slot, IoT devices receive a charge whenever the AP transmits. The receive signal power $p^t_{i}$ at IoT device $i$ is calculated as $p^t_{i}=g^t_{i}P_t$. We consider a practical $2.4$ GHz non-linear RF harvester~\cite{6920900}. For IoT device $i$, $\beta(p^t_{i})$ returns the RF-energy conversion rate of the harvester given incident power $p^t_i$.

Each IoT device consumes a fixed amount of energy, denoted as $\hat{E}$ (in mJ), to sample and return its data to the AP. If its energy level satisfies $b_i^t < \hat{E}$, device $i$ will not carry out any operation. Mathematically, $b_i^t$ evolves as follows,
\begin{equation}
\label{}
b_i^t=\left\{
\begin{aligned}
b_i^{t-1}+p^t_{i}\beta & , & b_i^t < \hat{E}, \\
\min(b_{max}, b_i^{t-1}+p^t_{i}\beta-\hat{E}) & , & \text{Otherwise}.
\end{aligned}
\right.
\end{equation}

In each time slot, the AP transmits to a random data user $u\in U$ that has a channel gain of $g^t_u$. Let $\gamma^t_u=p^t_{u}/N_0$ be the Signal-to-Noise Ratio (SNR) of user $u$, where $p^t_u$ is the received signal strength and $N_0$ is the white noise power.  
As per the Shannon-Hartley formula, its data rate is,
\begin{equation}
    r_u^t=W\log_2 (1+\gamma^t_u),
\end{equation}
where $W$ is the bandwidth of the channel.

Without loss of generality, we assume all users in $U$ have a fixed data rate requirement $r_{min}$.  
Define an indicator $J^t$ that returns one if $r_u^t \geq r_{min}$; otherwise, it returns zero.  Let $I^t$ be an indicator that returns a value of one if {\em all} IoT devices are able to collect a sample and transmit in slot $t$.

Our problem is as follows: find the optimal transmit power policy $\pi^*$ that returns the transmit power level for each time $t$ that maximizes,
\begin{equation}\label{obj-1}
Z_1=\lim_{T\rightarrow\infty} \frac{1}{t} \mathbb{E} \left[ \sum_{t=1}^T I^tJ^t \right].
\end{equation}

We will also consider an alternative aim, which is to maximize energy efficiency. That is,
\begin{equation}\label{energy efficiency}
\eta_t=\left\{
\begin{aligned}
\frac{I^tJ^t}{P_t} & , & P_t\not=0, \\
0 & , & P_t=0.
\end{aligned}
\right.
\end{equation}
In words, Eq.~\ref{energy efficiency} represents the fact that energy efficiency is higher if the AP uses less energy to support both types of users. In this respect, the second objective is to maximize,
\begin{equation}\label{obj-2}
Z_2=\lim_{T\rightarrow\infty} \frac{1}{t} \mathbb{E} \left[ \sum_{t=1}^T \eta_t \right].
\end{equation}

\section{Solutions}\label{Solution}
We first outline our RL solution followed by a solution that uses MPC. 

\subsection{Solution-1: Reinforcement Learning}
We first formulate our problem at hand as a Markov Decision Process (MDP)~\cite{BELLMAN1957MDP}, which is then solved using a Deep Q-network (DQN)~\cite{Mnih2015HumanlevelCT}.

\subsubsection{MDP}
Define a tuple with four elements $(S, A, P(s_{t+1}|s_t,a_t), R(s_{t+1}|s_t,a_t))$. The state space is $S$, where $s_t \in S$ represents the state at time $t$. The term $A$ represents the action space and $a_t \in A$ is the action taken at $t$. The transition probability to state $s_{t+1}$ after taking action $a_t$ is defined as $P(s_{t+1}\;|\;s_t,a_t)$. Lastly, the reward function $R(s_{t+1}|s_t,a_t)$ returns the immediate reward $r_{t}$ after taking action $a_t$ at state $s_t$. We assume there is an agent that observes the state, takes an action and then claims a reward. Let $\pi$ be the policy taken by an agent, where $\pi(s_t)$ returns the action $a_t$ when state $s_t$ is seen. Let $\pi^*$ be the optimal policy. The agent’s goal is to find the optimal policy $\pi^*$ that maximizes the following long-term cumulative reward,
\begin{equation}\label{long-term accumulative reward}
\mathop{\mathbb{E}}\left[\sum_{t=0}^{+\infty}R(s_{t+1}|s_t,\pi(s_t))\right].
\end{equation}

We are now ready to instantiate an MDP for our problem. Let the state $s_t$ be a tuple $s_t=(g^t_u,B_t)$ that includes the channel gain of a user and the battery level of the AP. 
The action $a_t$ corresponds to the transmission power level $P_t$ adopted by the AP at slot $t$. It has range $[0, P_{max}]$, meaning that the AP can choose a transmission power from $[0, P_{max}]$. Also, we discretize the action space equally to $N_A$ actions. In addition, if the AP's battery $B_t$ has insufficient energy to support the action chosen by the agent, then we encode the action to $a_t=B_t$. This means the AP consumes all the energy in its battery at slot $t$. 
As we consider two different objectives $Z_1$ and $Z_2$, which correspond to \eqref{obj-1} and \eqref{obj-2}, respectively, 
we define two reward definitions, denoted as \textbf{Reward-1} and \textbf{Reward-2}. 
For Reward-1, the reward $r_t$ for state $s_t$ and action $a_t$ is defined as $r_t=I^tJ^t$. To be specific, if the AP takes an action that is able to satisfy both types of users, then the reward is one; otherwise, it is zero. The second reward definition, namely Reward-2 relate to the energy efficiency $\eta_t$ achieved by the AP at slot $t$, and is defined as $r_t=\eta_t$. Reward-2 reveals the fact that the AP will claim more reward if it is able to achieve higher energy efficiency. Note, unless stated explicitly, our approach will apply Reward-1. The aforementioned transition of state-action-reward satisfies the Markov property. That is, the battery level in the next state $s_{t+1}$ depends on the current state battery level and the action taken at slot $t$. Also, the historical battery level and channel gain do not impact the transition of the current state $s_t$ to the next state $s_{t+1}$. To ensure our solution is practical, we assume the transition probability $P(s_{t+1}\;|\;s_t,a_t)$ is unknown. Hence, as our MDP is model-free.  To solve the formulated MDP, we will apply an RL approach, namely, Deep Q-network (DQN) to learn the optimal policy $\pi^*$ that maximizes \eqref{obj-1} or \eqref{obj-2}.

\subsubsection{DQN}
\begin{figure}[htbp]
      \centering
      \centerline{\includegraphics[width=0.5\textwidth]{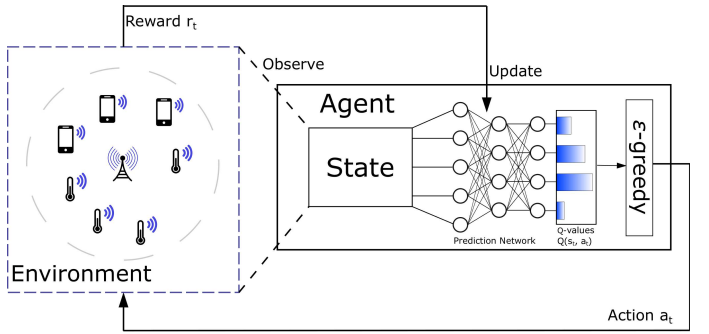}}
      \caption{An illustration of a DQN.}
      \label{DQN}
\end{figure}

The architecture of a DQN is shown in Figure~\ref{DQN}. Briefly, an agent located at the AP observes the environment (state) and takes an action based on its policy. The agent consists of a set of neural networks that outputs the expected value or Q-value of actions. It then observes the resulting reward, and uses it to update its policy. 

Let the Q-value $Q(s_t,a_t)$ denote the expected accumulated reward of a state-action after taking action $a_t$ at state $s_t$. A DQN uses a neural network to estimate Q-values. To do that, it includes two sets of neural networks; one of which denoted $\theta$ is for evaluation, and the other target network is denoted as $\theta'$.  A DQN aims to minimize the temporal difference-error of Q-values, defined as,
\begin{equation}\label{loss function}
L(\theta)=\min \displaystyle \mathop{\mathbb{E}}[(Q'(r_t,s_{t+1},a_{t+1},\theta')-Q(s_t,a_t,\theta))^2],
\end{equation} 
where $Q'(r_t,s_{t+1}$ is the targeted Q-value, given by the target network $\theta'$ and $Q(s_t,a_t,\theta)$ is the evaluated Q-value, provided by the evaluation network $\theta$. The training data is from a memory buffer that stores historical state-action-reward pairs. For every $K$ slots, we train our DQN using the Stochastic Gradient Descent method~\cite{DLnature} with a learning rate $\alpha$. For every $K'$ slots, where $K' \gg K$, we copy the evaluated network $\theta$ to become a new target network $\theta'$.

In each slot, the evaluation network $\theta$ will output the evaluated Q-value $Q(s_t,a_t,\theta)$ for each action. An agent then selects an action using the $\epsilon$-greedy policy. That is, it will take the action with the maximum Q-value with probability $(1-\epsilon_t)$, where $\epsilon$ is the exploration rate. Otherwise, a random action will be taken. To ensure the agent explores actions sufficiently, it uses a diminishing exploration rate, calculated as per,
\begin{equation}\label{epsilon function}
\epsilon_t=\max\left[1, \epsilon_{T}+\frac{\epsilon_{inc}}{(t+1)^2}\right],
\end{equation} 
where $\epsilon_0$ and $\epsilon_{T}$ are the initial and final exploration rate, respectively. The term $\epsilon_{inc}$ is the diminishing rate of $\epsilon_t$.

\subsection{Solution-2: MPC}
Also known as Receding Horizon Control (RHC), MPC is used to choose control actions over a moving time horizon. It relies on a prediction model to predict the system dynamics, e.g., prices, weather, heating requirements. The prediction is then used to build a virtual system model that simulates the changes of a real system several slot ahead. At slot $t$, MPC determines an optimal control action $u^*_t$ to maximize a performance index $V^*_t$ over a time window $[\tau, \tau+L]$. That is,
\begin{equation}\label{op function}
V_t^*=\max \frac{1}{L+1}\sum_{\tau=t}^{t+L}v_\tau(x_\tau,u_\tau),
\end{equation}
where $L$ is the number of predicted time slots. The term $x_\tau$ is the predicted system state, where $\tau$ represents a time slot in a virtual system model. Note, the system states $x_{\tau+1}, \dots, x_{\tau+L}$ are the system state given by a prediction model. The term $u_\tau$ denotes the action for predicted slot $\tau$. The instantaneous performance $v_\tau(x_\tau,u_\tau)$ depends on both the current state $x_\tau$ and control action $u_\tau$. Eq.~\ref{op function} is solved iteratively. 

In our case, the system state $x_t$ corresponds to the solar energy arrival at our AP $\widetilde{E}_t$ and channel gain of data users $g_u^t$ and IoT devices $\{g_i^t; i=1,\ldots,N\}$. 
The control action $u_t$ is the transmission power level taken by the AP, ranging from $[0, P_{max}]$. The performance index is defined as $v_\tau(.)= I^t J^t$. In addition, we also define $v_\tau(.)=\eta_t$ when we optimize the second objective $Z_2$, see Eq.~\eqref{obj-2}.

We employ Gaussian Process Regression (GPR)~\cite{williams2006gaussian} as our prediction model. A GPR model can be trained as a probabilistic nonparametric black-box to identify a non-linear dynamic system. Given a set of training data $\{(q_i,p_i);i=1,2, \ldots,k\}$, A GPR model is to learn the predictive response value $p'$ for a new input value $q'$. A GPR model is trained using the following linear regression model,
\begin{equation}\label{GPR}
p=q^Tw+\xi,
\end{equation}
where the coefficient $w$ and the error variance $\hat{\sigma}^2$ of $\xi \sim \mathcal{N}(0,\hat{\sigma}^2)$ are learned from training data. To build our training data, we let $q_i$ be the $i$-th slot and $p_i$ be the value of a network parameter; e.g., the energy arrival at the slot $i$.  For each network parameter, we employ an independent GPR model.  As an example, consider predicting the energy arrivals at the AP using GPR. We use set $D_t$ with size $k$ to store the energy arrivals in last $t$ slots; i.e., the set $D_t$ is a shifting window for energy arrivals $\{(i, \widetilde{E}_i);i=t-k,t-k-1,\ldots,t\}$ in the last $k$ slots. In each slot, we train a GPR model using the training data in set $D_t$, where the input value $p$ is time slot and the response value $q$ is the corresponding energy arrivals.  

\begin{figure}[htbp]
	\centering
	\centerline{\includegraphics[width=0.45\textwidth]{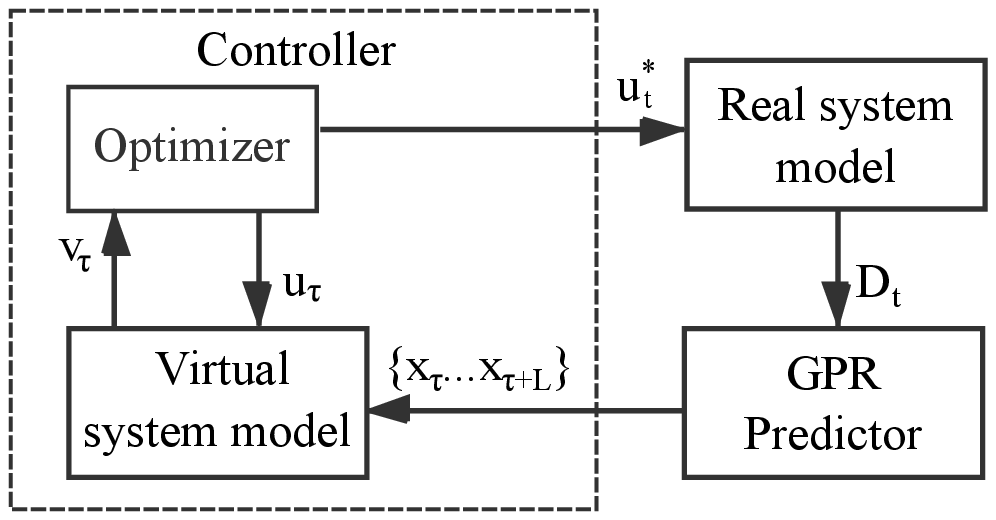}}
	\caption{An illustration of MPC.}
	\label{MPC model}
\end{figure}

Figure~\ref{MPC model} illustrates our MPC solution. First, it provides training data $D_t$ to the GPR predictor. At time slot $t$, the GPR predictor generates a set of system state predictions $x_{\tau+1} \dots x_{\tau+L}$, according to the current state $x_t$. With the predicted future states in hand, it builds a virtual system model that allows the optimizer to implement its virtual control action $u_\tau$. The optimizer we use to solve Eq.~\ref{op function} is binary search method with precision $\Psi$; see~\cite{binarysearch}. 
Then, the virtual system model evaluates the performance $v_\tau(.)$ given action $u_\tau$.   Once we have the optimal $V_t^*$, the optimizer will output the corresponding control action $u^*_t$ and apply $u^*_t$ in time slot $t$. Eq.~\ref{op function} is then solved iteratively until $t=T$.

%
\section{Evaluation}\label{Evaluation}
We conducted our simulations using Python 3.5 with TensorFlow 1.0 and Scikit-learn 0.21. Table~\ref{simparameters} shows all our parameter values. 
We {\em emphasize} that as our problem is new, there are no other solutions that we can benchmark against our solutions.  To this end, we benchmark DQN and MPC against the following algorithms/solutions/rules: 
\begin{itemize}
\item \textbf{Tabular RL (TRL)}: The state, action and reward are the same as that of DQN. However, the difference is that TRL uses a table to store Q-values and does not have the memory replay feature.
\item \textbf{Greedy}: In every slot, the AP uses $P_{max}$ to transmit; otherwise, it uses a transmission power level that consumes all the energy in its battery.
\item \textbf{Random}: The AP uniformly chooses a transmission power level in the range $[0, P_{max}]$.
\item \textbf{No-policy}: The AP is not aware of IoT devices and aims to only meet the requirement of data users. For a given data user $u$, the transmission power $P_t$ is calculated as 
\begin{equation}
P_t=\frac{N_0(2^{\frac{r_{min}}{W}}-1)}{g_u^t}.
\end{equation}
\end{itemize}

We assume each episode has 3,000 slots. For each episode, we calculate the average value of the following metrics: 
\begin{itemize}
\item \textbf{Number of activated IoT devices} per slot, which is calculated as $\frac{1}{3000}\sum^{t}_{t+3000}{n_t}$. This metric corresponds to the average number of IoT devices that achieve sampling per slot in one episode.
\item \textbf{Fraction of satisfied data users}. This is calculated as $\frac{1}{3000}\sum^{t}_{t+3000}{I_t}$, meaning the number of satisfied data users over one episode of slots.
\item \textbf{Energy efficiency}, which is equal to $\frac{1}{3000}\sum^{t}_{t+3000}{\eta_t}$.  This metric is the average energy efficiency achieved by an AP in one episode.
\item \textbf{Reward}. This is defined as the average reward gained by the agent within one episode; it is calculated as $\frac{1}{3000}\sum^{t}_{t+3000}{r_t}$, where ${r_t}$ refers to Reward-1.
\end{itemize}
%


%
\begin{table}[ht]	
    \caption{Parameter values used in our experiments.}\ 
	\label{simparameters}
	\begin{tabular}{ll}
		\\\midrule
		\textit{\textbf{Model Parameters}}                       & \textit{\textbf{Values}}      \\ \midrule
		Number of IoT Devices $N$                         & 5                   \\
		The minimum user distance $d_{min}$              & 5 meters                 \\
		The maximum user distance $d_{max}$                    & 25 meters                   \\
		The minimum device distance $\hat{d}_{min}$                          & 9 meters       \\
		The maximum device distance $\hat{d}_{max}$                       & 10 meters              \\
		The maximum battery of the AP $B_{max}$           & 100 J             \\
		The maximum battery of IoT devices $b_{max}$                  & 50 mJ     \\
		Solar panel size $\Phi$                   & 15 cm$^2$        \\
		Solar energy conversion efficiency $\overline{\eta}$       & 15\%                    \\
		The maximum transmission power of the AP $P_{max}$ & 200 mJ                  \\
		The mean of the Rayleigh distribution channel $\mu$  & 1                  \\
		The variance of the Rayleigh distribution channel $\sigma^2$         & 0.1   \\
		Bandwidth of the channel $W$       & 20 MHz      \\
		Energy requirement per sample $\hat{E}$  & 1.38 mJ \cite{7808981}   \\
		User data rate requirement $r_{min}$  & 133 Mbps   \\
		Total simulated time slots $T$          & 150,000 \\
		White noise power $N_0$          & $10^{-6}$ W \\
		\multicolumn{2}{l}{}                                             \\ \midrule
		\textit{\textbf{Algorithm Parameters}}                       & \textit{\textbf{Values}}      \\ \midrule
		Learning rate $\alpha$               & $10^{-5}$                 \\
		Reward decay rate $\gamma$                & 0.4                  \\
		Number of actions $N_A$             & 100          \\
		Memory size $N_{\mathcal{D}}$             & 50,000          \\
		Mini-batch size $N_{mb}$                   & 200                  \\
		Updating interval for $\theta$ $K$    & Every two slots     \\
		Replacing interval for $\theta'$ $K'$  & Every 400 slots      \\
		Replay start time slot                          & 3,000-th slot     \\
		Activation function                  & leaky ReLU              \\
		Initial exploration rate $\epsilon_0$     & 1                    \\
		Final exploration rate $\epsilon_{T}$     & 0.01                 \\
		Results collection time                   & the 120,000-th     \\
		Precision of MPC $\Psi$  & $10^{-31}$     \\
		Size of training data set $D_t$    & 20 \\
		GPR kernels & RBF \\
		Number of predicted slots by MPC $L$     & 4       \\ \hline
	\end{tabular}
\end{table}

\begin{figure}[htbp]
	\centering
	\centerline{\includegraphics[width=0.9\columnwidth]{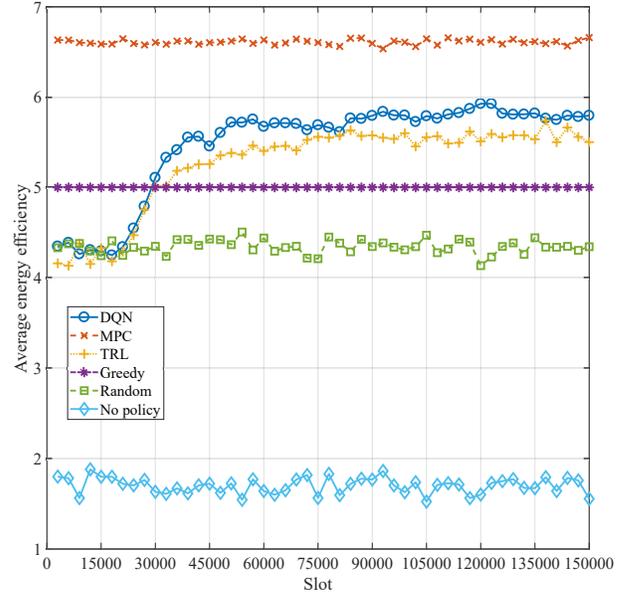}}
	\caption{Elapsed time versus energy efficiency.}
	\label{perfect_energy_EE}
\end{figure}

\begin{figure}[htbp]
	\centering
	\centerline{\includegraphics[width=0.9\columnwidth]{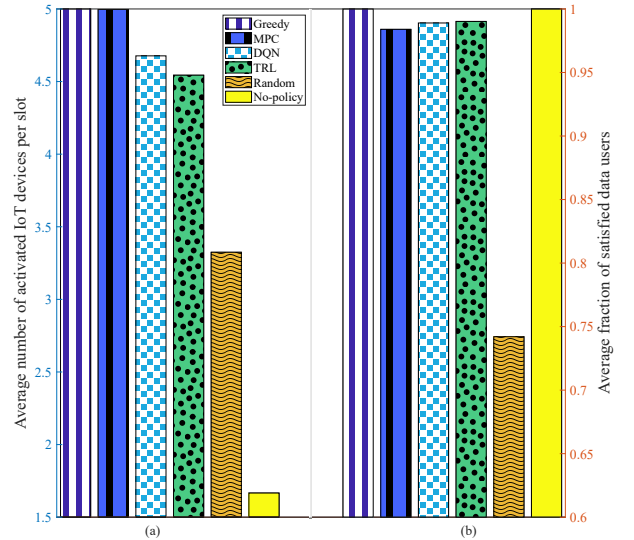}}
	\caption{The satisfaction of both types of users. (a) Average number of activated IoT devices per slot. (b) Average fraction of satisfied data users.}
	\label{perfect_energy_iot_n_user}
\end{figure}

We now investigate the energy efficiency $\eta_t$ of the AP assuming it has no energy constraint; e.g., when it is connected to the grid. We apply Reward-2, namely energy efficiency. 
From Figure~\ref{perfect_energy_EE}, we see that MPC achieves the highest energy efficiency of around 6.5.  In terms of DQN and TRL, we notice that energy efficiency increases because the RL agent learns to determine the optimal transmission power over time. Before the $15,000$-th slot, the agent has no knowledge of the environment. Hence, as we can see from Figure~\ref{perfect_energy_EE}, the energy efficiency achieved by both DQN and TRL agent is only 4.4. After training, the DQN gains the second-best energy efficiency among the tested algorithms, with around 5.8 on average. By contrast, TRL improves energy efficiency to only 5.5. Also, we see that the energy efficiency of TRL is worse than DQN. This is because DQN takes advantage of neural networks that allow it to deal with continuous and large state space. In our problem, the CSI is random and continuous values, so the state space is large.  Also, the memory replay strategy used by DQN breaks the correlation between adjacent states, which helps speed up convergence.

For non-RL approaches, such as Greedy, Random and No-policy, the energy efficiency shown in Figure~\ref{perfect_energy_EE} remains roughly the same. For example, the energy efficiency achieved by Random is always 4.4. This is because the transmission power used by Random is uniformly distributed in the range $[0, 0.2]$ mW, meaning that the average energy consumption per slot/transmission power is 0.1 mW. Such a transmission power level is not sufficient to meet the energy/data rate demands of users.
In terms of No-policy, the energy efficiency is around 1.8 because the transmission power used by No-policy is not sufficient to activate every IoT device. As we can see from Figure~\ref{perfect_energy_iot_n_user}, the number of activated IoT devices is only 1.7 

From Figure~\ref{perfect_energy_iot_n_user}, we see that Greedy has the highest satisfaction for both IoT and legacy data users, with five active devices per slot with a user satisfaction of 100\%, respectively. 
This is because Greedy is able to use the maximum transmit power (200 mW) because there is no energy limitation in this scenario. MPC also achieves almost 100\% user satisfaction with five active devices per slot. In terms of data users, MPC achieves a satisfaction value of 0.98. However, from Figure~\ref{perfect_energy_EE}, we see that energy efficiency achieved by Greedy is always 5.0, which is lower than MPC by 35\%. This means MPC uses 35\% less energy to achieve almost the same performance as Greedy. As for DQN, Figure~\ref{perfect_energy_iot_n_user} shows that the average number of activated devices per slot is 4.65 and the satisfaction of data users is 0.98. 
This indicates that MPC performs better than DQN in terms of supporting IoT devices. 
Also, Figure~\ref{perfect_energy_EE} and~\ref{perfect_energy_iot_n_user} indicate that compared to Greedy, DQN support 6\% less number of IoT devices per slot but its energy efficiency is 16\% higher than Greedy.
Lastly, we see that No-policy gains 100\% user satisfaction but it only supports 1.7 IoT devices per slot on average. These results confirm that our solutions are able to effectively charge RF-energy IoT devices.

\begin{figure}[htbp]
	\centering
	\centerline{\includegraphics[width=0.9\columnwidth]{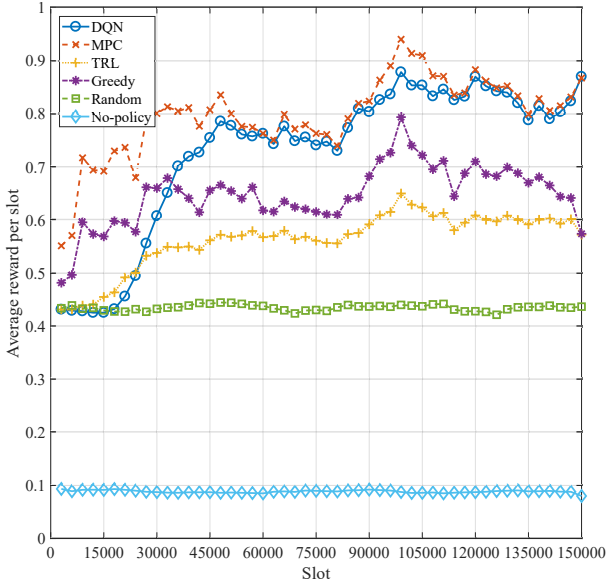}}
	\caption{Elapsed time versus reward gained by the tested algorithms/schemes.}
	\label{imperfect_energy_reward}
\end{figure}

\begin{figure}[htbp]
	\centering
	\centerline{\includegraphics[width=0.47\textwidth]{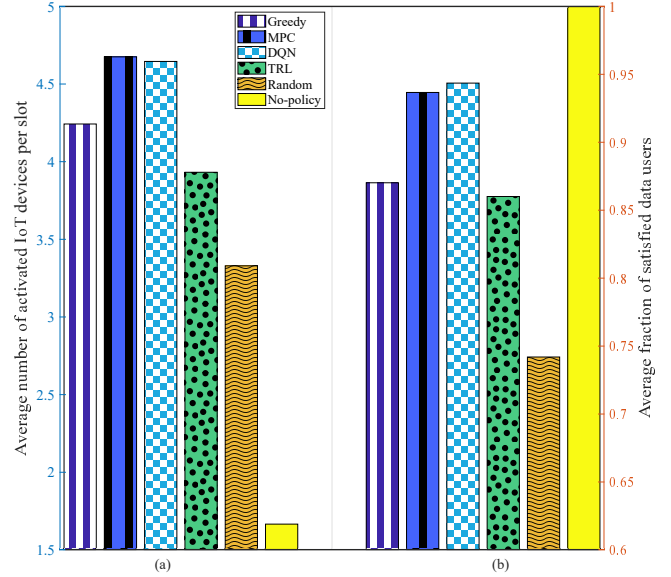}}
	\caption{The satisfaction of both types of users under a random energy arrival scenario. (a) Average number of activated IoT devices per slot, and (b) Average fraction of satisfied data users.}
	\label{imperfect_energy_iot_n_user}
\end{figure}

Next, we consider an imperfect energy supply. We set the solar panel size $\Phi$ to 15 cm$^2$.  The energy arrival at the AP is random. Figure~\ref{imperfect_energy_reward} shows the satisfaction of both data users and RF-energy devices achieved by the tested algorithms over 150,000 time slots. 
From Figure~\ref{imperfect_energy_reward}, we notice that the reward increases significantly because the RL agent learns to use energy over time. 
Before the $15,000$-th slot, the average reward gained by both DQN and TRL agent is only 0.43. After training, the DQN agent gains around 0.8 rewards on average, while TRL improves the reward to only 0.6. Also, we see from Figure~\ref{imperfect_energy_reward} that DQN gains around 20\% to 30\% more reward than TRL. As a result, Figure~\ref{imperfect_energy_iot_n_user} shows that DQN is able to support 20\% more IoT devices than TRL, where the number of activated IoT devices achieved by DQN and TRL is 4.61 and 3.9, respectively.

In terms of non-RL algorithms, Figure~\ref{imperfect_energy_reward} shows that MPC gains a reward between 0.75 and 0.9 which is the same as the well-trained DQN agent after the $60,000$-th slot.  MPC also supports 4.65 IoT devices and 0.94 data users per slot on average. 
We also see that Greedy gains around 0.65 reward on average.  However, its performance is worst than the case when the AP has no energy limitation.  This is because it does not conserve energy, meaning it causes energy outage. As for the Random rule, from Figure~\ref{imperfect_energy_reward}, it only gains 0.44 reward on average. This is because Random only uses $0.1$ mW energy per slot, leading to battery overflow in around 90\% time slots. 
No-policy gains only 0.1 reward; see Figure~\ref{imperfect_energy_reward}. 
The reason is that the number of activated IoT devices achieved by No-policy is only 1.7 per slot on average, meaning IoT devices require more slots to harvest energy until they have sufficient energy to gather a sample.
%
It is worth noting that the performance of DQN and MPC is not significantly affected by random energy arrivals.  
Comparing Figure~\ref{perfect_energy_iot_n_user} and ~\ref{imperfect_energy_iot_n_user}, when the AP has imperfect information of its energy arrivals, the user satisfaction of DQN and MPC reduced by around 8\% to 4\% for both types of users.
By contrast, we see from Figure~\ref{perfect_energy_iot_n_user} and ~\ref{imperfect_energy_iot_n_user} that the performance achieved by TRL, Greedy and Random drops significantly when we consider random energy arrivals. 
For example, the average number of activated IoT devices achieved by TRL is 3.9 per slot, whereas it is 4.55 when the AP is powered by a perfect energy source. The reason is that the state space must also include different energy levels at the AP and CSI of users, so the state space is larger.
In the case of random energy arrivals, the shortcomings of TRL when for large state space is evident. Moreover, our results show that when we consider an imperfect energy source, transmit power control is critical. The reason is because the AP may encounter energy shortage, which reduces system performance.  This is evident in Figure~\ref{perfect_energy_iot_n_user} and ~\ref{imperfect_energy_iot_n_user}, the user satisfaction of Greedy and TRL reduces by around 11\% to 15\%.

\begin{figure}[htbp]
	\centering
	\centerline{\includegraphics[width=0.9\columnwidth]{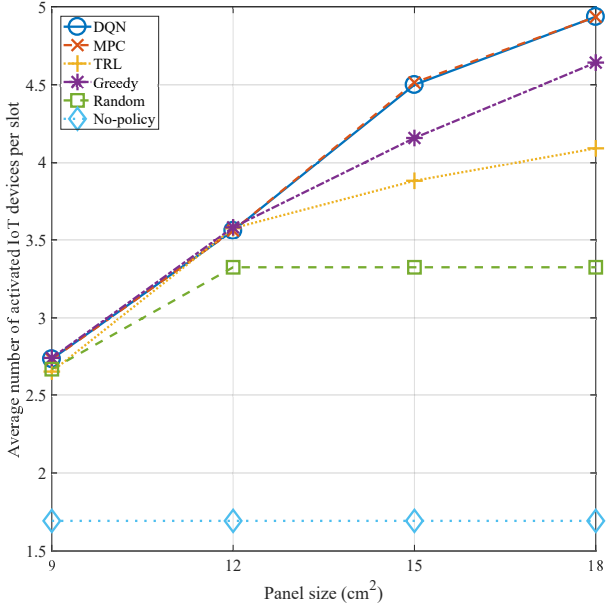}}
	\caption{Varying solar panel sizes versus the number of activated devices.}
	\label{panel_size_iot}
\end{figure}

\begin{figure}[htbp]
	\centering
	\centerline{\includegraphics[width=0.9\columnwidth]{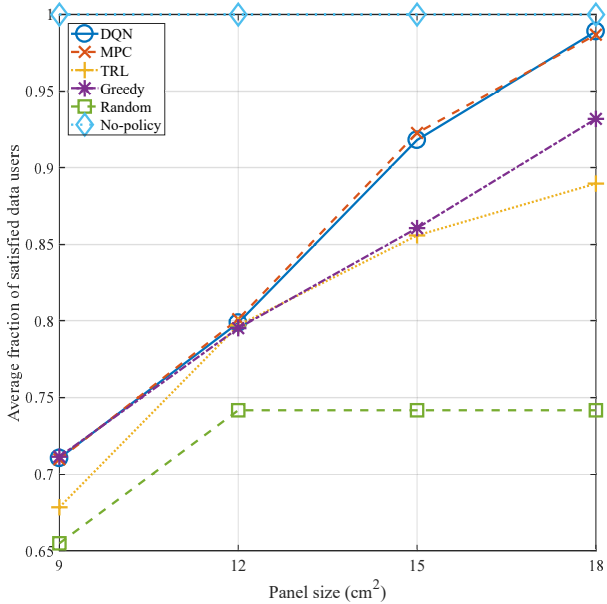}}
	\caption{Varying solar panel sizes versus the fraction of satisfied data users.}
	\label{panel_size_user}
\end{figure}

Figure~\ref{panel_size_iot} and~\ref{panel_size_user} illustrate how different solar panel sizes $\Phi$ impact user satisfaction. From Figure~\ref{panel_size_iot} and~\ref{panel_size_user}, we see that the satisfaction of both types of users increases as the panel size increases. The average number of activated IoT devices of DQN and RMCP increases from 2.75 to 4.9, with a 90\% increment as the panel size doubles. Also, the performance of data users increases by 40\%, from 0.71 to 0.98. This is because the AP harvests more energy when using a larger solar panel. This allows them to use a higher transmission power to meet the needs of those users that are far away from the AP.
In terms of Greedy and TRL, both of them are able to support  3.55 IoT devices per slot on average, and achieve 80\% satisfaction for data users when the solar panel size $\Phi$ is 12 cm$^2$, see Figure~\ref{panel_size_iot} and~\ref{panel_size_user}. However, they are inferior to DQN and MPC when $\Phi$ is larger than 12 cm$^2$. As shown in Figure~\ref{panel_size_iot} and~\ref{panel_size_user}, the number of activated IoT devices achieved by Random is always 3.3 per slot when the solar panel size is larger than 12 cm$^{2}$. This is because its average energy consumption is 0.1 mW. Moreover, as the solar panel size increases, higher energy arrivals lead to a higher overflow rate. 
As shown in Figure~\ref{panel_size_iot} and~\ref{panel_size_user}, the satisfaction of both data users and IoT devices gained by No-policy also remains unchanged as the solar panel size $\Phi$ increases. This indicates that the overflow rate of No-policy is higher than other solutions. The number of activated IoT devices is only 1.7. The reason is that the AP is not aware of IoT devices and thus those devices cannot harvest enough energy. 
Figure~\ref{panel_size_iot} and~\ref{panel_size_user} also show that the difference in user satisfaction achieved by different algorithms/schemes is wider as the panel size increases. 
This means energy management is more necessary when the solar energy arrival rate is large. 
This is because when the panel size is smaller than 12 cm$^2$, the AP is not able to harvest sufficient energy to meet the data/energy requirements of users. By contrast, our DQN and MPC algorithms are able to learn the optimal transmission power to maximize user satisfaction. Therefore, we see that our approaches are significantly superior to the other tested algorithms when the panel size is larger than 15 cm$^2$

\begin{figure}[htbp]
	\centering
	\centerline{\includegraphics[width=0.9\columnwidth]{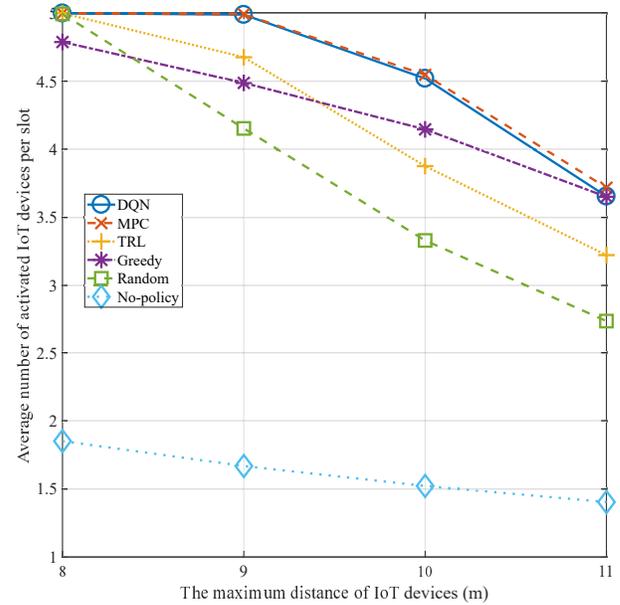}}
	\caption{Average device distance versus the number of activated devices.}
	\label{cell_size_iot}
\end{figure}

\begin{figure}[htbp]
	\centering  
	\centerline{\includegraphics[width=0.9\columnwidth]{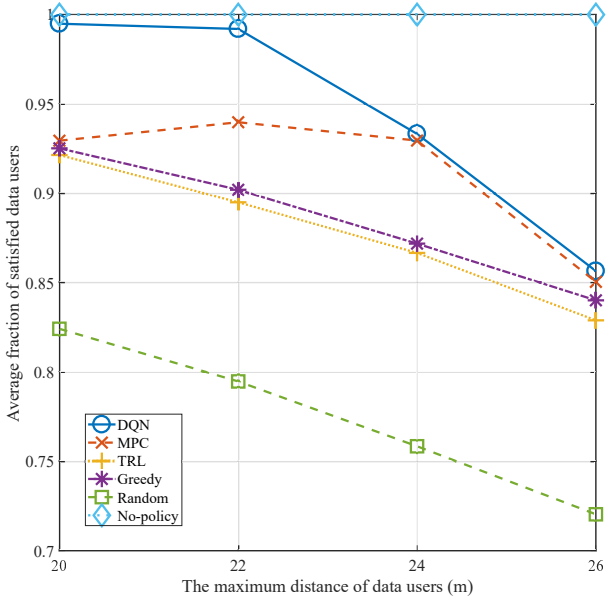}}
	\caption{Maximum user distance versus the fraction of satisfied data users.}
	\label{cell_size_user}
\end{figure}

We now study different cell sizes. Figure~\ref{cell_size_iot} and ~\ref{cell_size_user} show that the satisfaction of both types of users decreases by 35\% to 45\% as the distance/cell size increases. This is because the channel gain becomes smaller if the user distance increases. However, we see that our proposed DQN has minimal degradation in user satisfaction as the cell size becomes larger. For example, the fraction of satisfied data users decreases from 0.98 to 0.87, an 11\% drop as the IoT devices distance increases from 20 to 26 meters. As shown in Figure~\ref{cell_size_user}, our DQN algorithm is able to improve the percentage of satisfied data users by 8\% to 10\% when the maximum user distance is under 24 meters. 
Figure~\ref{cell_size_iot} shows that MPC achieves the same number of activated IoT devices to DQN as the maximum IoT device distance increases from eight to ten meters.   However, referring to Figure~\ref{cell_size_user}, MPC is no better than DQN in terms of the fraction of satisfied data users, which is the second-best algorithms among the tested algorithms.
Figure~\ref{cell_size_user} also shows that as the cell size increases, No-policy supports less RF-energy IoT devices. It only supports 1.4 devices per slot when the maximum IoT device distance is 11 meters. However, the user satisfaction of data users achieved by No-policy is unchanged; see Figure~\ref{cell_size_user}. The reason is that No-policy is only aware of data users so it will not increase transmission power if there is energy shortage at IoT devices.

We see from Figure~\ref{cell_size_iot} and~\ref{cell_size_user} that the user satisfaction achieved by different tested algorithms converges to a smaller value as user distance increases. 
For example, in Figure~\ref{cell_size_user}, the difference in the fraction of satisfied data users achieved by DQN and MPC converges to 0.01 as the maximum distance of data users increases to 26 meters. 
This reveals that the AP cannot gain high user satisfaction though transmission power management as the distance increases. The reason is that the channel gain becomes smaller with increasing user/device distance. 
Consequently, the received power at devices/users is too small for data receptions or energy harvesting no matter what transmission power policy is adopted by the AP. 
%

\section{Conclusion}\label{CONC}
We have shown how an AP can learn to adapt its transmit power when serving data users and simultaneously ensure IoT devices receive sufficient energy.  Our work is significant because existing networks will likely be used to support RF-energy harvesting IoT devices. Our results show that the proposed DQN and MPC algorithms power 10\% to 42\% more IoT devices and gain 9\% to 27\% more user satisfaction as compared to approaches without learning capabilities. Also, the results show that our algorithms are able to determine the optimal transmission power for devices and users according to the current and historical battery status.
%

\bibliographystyle{IEEEtran}
\bibliography{allrefs}

\end{document}